\documentclass[prd,aps,nofootinbib,onecolumn]{revtex4}
\usepackage{amsmath}
\usepackage{epsfig,graphicx}
\usepackage{amssymb}
\usepackage{color}

\def\be{\begin{eqnarray}}
\def\ed{\end{eqnarray}}
\def\non{\nonumber}

\def\ga{\gamma}
\def\la{\langle}
\def\ra{\rangle}

\begin{document}

\title{\bf Branching ratios and direct CP asymmetries in $D\to PP$ decays}

\author{Hsiang-nan Li$^{1,2,3}$, Cai-Dian Lu$^4$, Fu-Sheng Yu$^{4,5}$}
\affiliation{$^1$ Institute of Physics, Academia Sinica, Taipei,
Taiwan 115, Republic of China,\\
$^{2}$Department of Physics, National Cheng-Kung University, Tainan,
Taiwan 701, Republic of China \\
$^{3}$Department of Physics, National Tsing-Hua university,
Hsin-Chu, Taiwan 300, Republic of China \\
$^{4}$Institute of High Energy Physics and Theoretical Physics
Center for Science  Facilities, Chinese Academy of Sciences, Beijing
100049, People's Republic of China \\
 $^5$  Laboratoire de
l'Acc$\acute{e}$l$\acute{e}$rateur Lin$\acute{e}$aire,
    Universit$\acute{e}$ Paris-Sud 11, CNRS/IN2P3(UMR 8607) 91405
    Orsay, France}

\begin{abstract}

We propose a theoretical framework for analyzing two-body
nonleptonic $D$ meson decays, based on the factorization of
short-distance (long-distance) dynamics into Wilson coefficients
(hadronic matrix elements of four-fermion operators). The
parametrization of hadronic matrix elements in terms of several
nonperturbative quantities is demonstrated for the $D\to PP$ decays,
$P$ denoting a pseudoscalar meson. We consider the evolution of
Wilson coefficients with energy release in individual decay modes,
and the Glauber strong phase associated with the pion in
nonfactorizable annihilation amplitudes, that is attributed to the
unique role of the pion as a Nambu-Goldstone boson and a
quark-anti-quark bound state simultaneously. The above inputs
improve the global fit to the branching ratios involving the $\eta'$
meson, and resolves the long-standing puzzle from the
$D^0\to\pi^+\pi^-$ and $D^0\to K^+K^-$ branching ratios,
respectively. Combining short-distance dynamics associated with
penguin operators and the hadronic parameters determined from the
global fit to branching ratios, we predict direct CP asymmetries, to
which the quark loops and the scalar penguin annihilation give
dominant contributions. In particular, we predict $\Delta A_{\rm
CP}\equiv A_{\rm CP}(K^+K^-)-A_{\rm CP}(\pi^+\pi^-)=-1.00\times
10^{-3}$, lower than the LHCb and CDF data.
\end{abstract}

\maketitle %

\section{Introduction}

Two-body nonleptonic $D$ meson decays have attracted great attention
recently. The very different data for the $D^0\to\pi^+\pi^-$ and
$D^0\to K^+K^-$ branching ratios have been a long-standing puzzle:
the former (latter) is lower (higher) than theoretical predictions
from analyses based on the topology parametrization
\cite{BR10,diag}. The deviation in these two $D\to PP$ modes, $P$
representing a pseudoscalar meson, stands even after taking into
account flavor $SU(3)$ symmetry breaking effects in emission
amplitudes \cite{diag}. Another puzzle appears in the difference
between the direct CP asymmetries of the $D^0\to K^+ K^-$ and
$D^0\to \pi^+\pi^-$ decays,
 \be
 \Delta
 A_{\rm CP}\equiv A_{\rm CP}(K^+K^-)-A_{\rm CP}(\pi^+\pi^-)=
 [-0.82\pm0.21({\rm stat})\pm0.11({\rm syst})]\%,\label{lhcb}
 \ed
measured by LHCb collaboration \cite{arXiv:1112.0938}, which is the
first evidence of CP violation in charmed meson decays. The above
result has been confirmed by the CDF collaboration
\cite{CDF12}\footnote{The CDF collaboration observed the smaller
direct CP asymmetries $A_{\rm CP} (D^0 \to\pi^+\pi^-) = (+0.22 \pm
0.24 \pm 0.11)\%$ and $A_{\rm CP} (D^0 \to K^+K^-) = (-0.24 \pm 0.22
\pm 0.09)\%$ in \cite{Aaltonen:2011se}.},
 \be
 \Delta
 A_{\rm CP}=
 [-0.62\pm0.21({\rm stat})\pm0.10({\rm syst})]\%.\label{cdfn}
 \ed
The quantity $\Delta A_{\rm CP}$ is naively expected to be much
smaller in the Standard Model (SM), since the responsible penguin
contributions are suppressed by both the CKM matrix elements and the
Wilson coefficients \cite{Grossman:2006jg,Bigi:2011re},
\begin{eqnarray}
A_{\rm CP}\sim {|V_{cb}^*V_{ub}|\over |V_{cs}^*V_{us}|}
{\alpha_s\over\pi}\sim10^{-4}.
\end{eqnarray}

The above two puzzles have triggered intensive investigations in the
literature employing different approaches. The mechanism responsible
for the very different $D^0\to\pi^+\pi^-$ and $D^0\to K^+K^-$
branching ratios is still unclear. It has been speculated that the
long-distance resonant contribution through the nearby resonance
$f_0(1710)$ in the $W$-exchange topology could explain why a $D^0$
meson decays more abundantly to the $K^+K^-$ than $\pi^+\pi^-$ final
state \cite{diag}: it is attributed to the dominance of the scalar
glueball content in $f_0(1710)$ and the chiral suppression on the
scalar glueball decay into two pseudoscalar mesons. This mechanism
was systematically formulated in the single pole-dominance model
with additional arbitrary ``Wilson coefficients", and then included
into the topology parametrization for global fits
\cite{Fusheng:2011tw}. Unfortunately, the predicted
$D^0\to\pi^+\pi^-$ ($D^0\to K^+K^-$) branching ratio remains higher
(lower) than the data. That is, a systematic and global
understanding of the $D\to PP$ decays has not existed yet. Because
of the high precision of the two data, even a small deviation
deteriorates the global fit. If the branching ratios are not well
explained, some important decay mechanism may be still missing, and
any predictions for the direct CP asymmetries are not convincing.

For the second puzzle, a reliable evaluation of the penguin
contributions to two-body nonleptonic $D$ meson decays is not
available so far. In \cite{Cheng:2012wr,Brod:2011re} the tree
amplitudes were determined by fitting the topology parametrization
to the data of the branching ratios, while the penguin amplitudes
were calculated in the QCD-improved factorization (QCDF)
\cite{BBNS99,BBNS01}. That is, the tree and penguin contributions
were treated in the different theoretical frameworks. It has been
noticed that the penguin amplitudes derived from QCDF lead to a tiny
$\Delta A_{\rm CP}$ of order $10^{-5}$ \cite{Cheng:2012wr}. Allowing
the penguin amplitudes to be of the same order as the tree ones
discretionally, $\Delta A_{\rm CP}$ reaches $-0.13\%\sim O(10^{-3})$
\cite{Cheng:2012wr}. In another work \cite{Bhattacharya:2012ah} also
based on the topology parametrization, the penguin contribution via
an internal $b$ quark was identified as the major source of CP
violation, since it cannot be related to the tree amplitudes. This
penguin contribution, including its strong phase, was constrained by
the LHCb data, and then adopted to predict direct CP asymmetries of
other decay modes. In this way it is difficult to tell whether the
large $\Delta A_{\rm CP}\sim O(10^{-2})$ in Eqs.~(\ref{lhcb}) and
(\ref{cdfn}) arise from new physics
\cite{Feldmann:2012js,Chen:2012am,Altmannshofer:2012ur,Giudice:2012qq,
Hochberg:2011ru,Isidori:2011qw,Rozanov:2011gj}. The similar comment
applies to the analysis in \cite{Pirtskhalava:2011va}. Viewing that
the postulated $\Delta A_{\rm CP}$ ranges from $10^{-5}$ to
$10^{-2}$ in the SM \cite{Franco:2012ck}, it is crucial to give precise
predictions for the direct CP asymmetries in the $D\to PP$ decays.

In this paper we aim at proposing a theoretical framework, in which
various contributions, such as emission and annihilation ones, and
tree and penguin ones are all handled consistently. The only
assumption involved is the factorization of short-distance dynamics
and long-distance dynamics in two-body nonleptonic $D$ meson decays
into Wilson coefficients and hadronic matrix elements of
four-fermion operators, respectively. For the hadronic matrix
elements, including the emission, $W$-annihilation, and $W$-exchange
amplitudes, the treatment is similar to the topology parametrization
in the factorization hypothesis. The quark-loop and magnetic penguin
contributions are absorbed into the Wilson coefficients for the
penguin operators. As to the scale of the Wilson coefficients, we
set it to the energy release in individual decay modes, which
depends on masses of final states. An important ingredient is the
Glauber strong phase factor \cite{CL11} associated with a pion,
whose strength has been constrained by the data of the $B\to \pi K$
direct CP asymmetries \cite{Li:2009wba}. Briefly speaking, we have
improved the topology parametrization by taking into account
mode-dependent QCD dynamics, for instance, flavor $SU(3)$ symmetry
breaking effects, in two-body nonleptonic $D$ meson decays.

Due to the small charm quark mass $m_c\approx1.3$ GeV, a
perturbative theory may not be valid for $D$ meson decays. However,
with the abundant data, all the above hadronic parameters can be
determined. It will be shown that the Glauber phase associated with
a pion is crucial for understanding the $D^0\to\pi^+\pi^-$ branching
ratio very different from the $D^0\to K^+K^-$ one. This additional
phase modifies the relative angle and the interference between the
emission and annihilation amplitudes involving pions. The predicted
$D^0\to\pi^+\pi^-$ branching ratio is then reduced, while the
predicted $D^+\to\pi^+\eta$ branching ratio is enhanced, such that
the overall agreement with the data is greatly improved. Our scale
choice for the Wilson coefficients impacts the $D\to PP$ decays
containing the $\eta'$ meson in the final states: the $\eta'$ meson
mass about 1 GeV is not negligible compared to the $D$ meson mass,
so the energy release in these modes should be lower. Once the
hadronic parameters are fixed by the measured branching ratios, the
penguin amplitudes, expressed as the combination of the hadronic
parameters and the corresponding Wilson coefficients, are also fixed
in our framework. That is, we can predict the direct CP asymmetries
of the $D\to PP$ decays in the SM without ambiguity. In particular, we
obtain $\Delta A_{\rm CP}=-1.00\times 10^{-3}$, which discriminates
the opposite postulations on large (small) direct CP asymmetries in
singly Cabibbo-suppressed $D$ meson decays \cite{GG89}
(\cite{pietro}). It will be observed that the quark loops and the
scalar penguin annihilation generate large relative phases between
the tree and penguin amplitudes, and are the dominant sources of the
direct CP asymmetries.

In Sec.~II we present our parametrization of the tree contributions
to the $D\to PP$ branching ratios based on the factorization
hypothesis with QCD dynamics being implemented. In Sec.~III the
penguin contributions from the operators $O_{3-6}$, from $O_{1,2}$
through the quark loops, and from the magnetic penguin $O_{8g}$ are
formulated. The direct CP asymmetries in the $D\to PP$ decays are
then predicted. Section~IV is the conclusion. We show the explicit
topology parametrization for the $D^0\to\pi^+\pi^-$ and $D^0\to
K^+K^-$ decays in Appendix~\ref{app:amp}. Appendix~\ref{app:wc}
collects the evolution formulas for the Wilson coefficients in the
scale $\mu$. The $\eta$-$\eta'$ mixing formalism and the chiral
factors of their flavor states are given in Appendix~\ref{anomaly}.
The form factors involved in the factorizable scalar penguin
annihilation are specified in Appendix~\ref{me}.

\section{Branching Ratios}

\begin{figure}
\begin{center}
\includegraphics[scale=0.5]{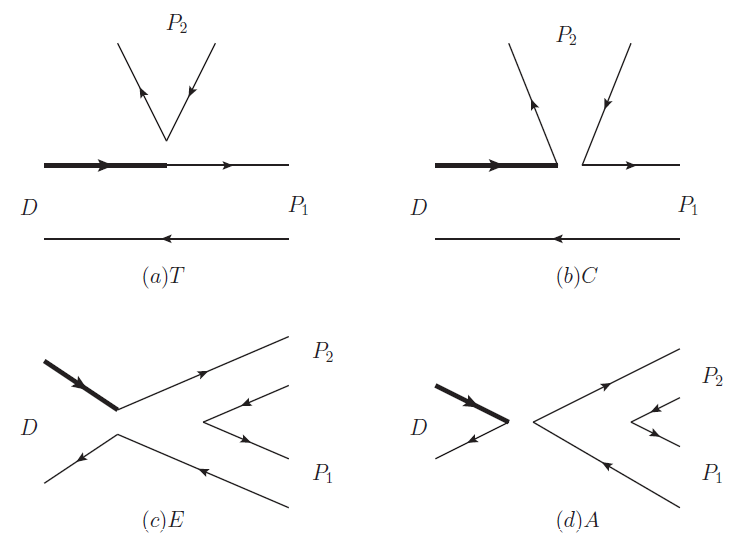}
\end{center}
\caption{Four topological diagrams contributing to $D\to PP$ decays:
(a) the color-favored tree amplitude $T$, (b) the color-suppressed
tree amplitude $C$, (c) the $W$-exchange amplitude $E$, and (d) the
$W$-annihilation amplitude $A$. The thick line represents the charm
quark. }\label{fig:topo}
\end{figure}

In this section we formulate the tree contributions, which dominate
the branching ratios of charm decays. The relevant weak effective
Hamiltonian is written as
\begin{equation}\label{Heff12}
\mathcal{H}_{eff}=\frac{G_F}{\sqrt{2}}
 V_{CKM}\left[C_1(\mu)O_1(\mu)+C_2(\mu)O_2(\mu)\right]+h.c.,
\end{equation}
where $G_F$ is the Fermi coupling constant,
$V_{CKM}$ denotes the corresponding Cabibbo-Kobayashi-Maskawa
(CKM) matrix elements, $C_{1,2}$ are the Wilson coefficients, and
the current-current operators are
\begin{eqnarray}
& &O_1=(\bar{u}_{\alpha}q_{2\beta})_{V-A}
(\bar{q}_{1\beta}c_{\alpha})_{V-A},\nonumber\\
& &O_2=(\bar{u}_{\alpha}q_{2\alpha})_{V-A}
(\bar{q}_{1\beta}c_{\beta})_{V-A},
\end{eqnarray}
with $\alpha,\beta$ being color indices, and $q_{1,2}$ being the $d$
or $s$ quark. There are four types of topologies dominantly
contributing to the branching ratios of $D\to P_1P_2$ decays
\cite{topo}, where $P_2$ represents the meson emitted from the weak
vertex: the color-favored tree amplitude $T$, the color-suppressed
amplitude $C$, the $W$-exchange amplitude $E$, and the
$W$-annihilation amplitude $A$, shown in Fig.~\ref{fig:topo}.
Penguin contributions are neglected for branching
ratios due to the suppression of the CKM matrix elements
$V_{cb}^*V_{ub}$. Besides, there exist flavor-singlet amplitudes,
in which a quark-antiquark pair produced from vacuum generates
color- and flavor-singlet states like $\eta^{(\prime)}$. However,
they are also neglected in this work, since the current
data do not suggest significant flavor-singlet contributions
\cite{diag,BR10}.

It is known that the heavy quark expansion does
not work well for charm decays because of the small $m_c$, so that
nonfactorizable contributions, such as vertex corrections,
final-state interaction (FSI), resonance effects, etc, cannot be
ignored, especially in the amplitude $C$. We parameterize these
contributions into a factor $\chi_{nf}$ . For simplicity, we drop
nonfactorizable contributions in the amplitude $T$, which is
dominated by factorizable ones. In general, there exists a relative
strong phase $\phi$ between $T$ and $C$, which may arise from
inelastic FSI or other long-distance dynamics in $D$ meson decays
\cite{Fusheng:2011tw,diag}. Therefore, the emission amplitudes $T$
and $C$ are parameterized as
\begin{eqnarray}\label{emit}
\langle P_1P_2|\mathcal{H}_{eff}|D\rangle_{T,C}
&=&\frac{G_F}{\sqrt{2}}V_{CKM}
a_{1,2}(\mu)f_{P_2}(m_D^2-m_{P_1}^2)F_0^{DP_1}(m_{P_2}^2),
\end{eqnarray}
with the $P_1$ ($P_2$) meson mass $m_{P_1}$ ($m_{P_2}$), the $P_2$
meson decay constant $f_{P_2}$, and the $D\to P_1$ transition form
factor $F_0^{DP_1}$. The scale-dependent Wilson coefficients are
given by
\begin{eqnarray}
& &a_1(\mu)=C_2(\mu)+{C_1(\mu)\over N_c}, \nonumber\\
& &a_2(\mu)=C_1(\mu)+C_2(\mu)\left[{1\over
N_c}+\chi_{nf}e^{i\phi}\right],
\end{eqnarray}
for $T$ and $C$, respectively, with $N_c=3$ being the number of
colors.

The flavor $SU(3)$ symmetry breaking effects have been known to be
significant in the $D\to PP$ decays, especially in the singly
Cabibbo-suppressed modes \cite{Wu:2004ht,Fusheng:2011tw,diag}. In
additional to the different decay constants $f_{P_2}$ and form
factors $F_0^{DP_1}$, the mass ratios $m_{K,\eta^{(\prime)}}/m_D$
should be distinguished too, with $m_K$, $m_{\eta^{(\prime)}}$, and
$m_D$ being the masses of the kaon, the $\eta^{(\prime)}$ meson, and
the $D$ meson, respectively. As suggested in the perturbative QCD
(PQCD) approach \cite{pqcd}, the scale $\mu$ is set to the energy
release in individual decay processes, which depends on final-state
masses. We propose the choice
 \be
 \mu=\sqrt{\Lambda~ m_D (1-r_2^2)},\label{scale}
 \ed
where $r_2=m_{P_2}^2/m_D^2$ is the mass ratio of the $P_2$ meson
over the $D$ meson, and $\Lambda$, representing the momentum of soft
degrees of freedom in the $D$ meson, is a free parameter. The
evolution of the Wilson coefficients in the scale $\mu$ is shown in
Appendix \ref{app:wc}. The above choice further takes into account
the $SU(3)$ breaking effect in the Wilson coefficients for the
emission amplitudes. It is then legitimate to regard $\chi_{nf}$,
$\phi$ and $\Lambda$ as being universal to all modes, which will be
determined by experimental data.

The $D^0\to K^0\overline K^0$ branching ratio through the
pure $W$-exchange channel vanishes in the $SU(3)$ limit due to the
cancelation of the CKM matrix elements \cite{diag,Eeg:2001un}. Hence, the large
observed branching ratio of this mode manifests the breaking of the
flavor $SU(3)$ symmetry in the $W$-exchange contribution, which
becomes almost of the same order as the emission one. Both the
$W$-exchange topology for neutral $D$ meson decays and the
$W$-annihilation topology for charged $D$ meson decays will be
included in our framework. Since the factorizable contributions to
these amplitudes are down by the helicity suppression, we consider
only the nonfactorizable contributions. Then the former (latter) is
proportional to the Wilson coefficient $C_2$ ($C_1$). Based on the
above reasoning, we parameterize the amplitudes $E$ and $A$ as
\begin{eqnarray}\label{ann}
\la P_1P_2|\mathcal{H}_{eff}|D\ra_{E,A}={G_F\over\sqrt{2}}V_{CKM}
b^{E,A}_{q,s}(\mu)f_{D} m_D^2  \left({f_{P_1}f_{P_2}\over
f_\pi^2}\right),
\end{eqnarray}
with the factors
 \be\label{b}
& &b_{q,s}^E(\mu)= {C_2(\mu)} \chi_{q,s}^E e^{i\phi_{q,s}^E},
  \nonumber\\
& &b_{q,s}^A(\mu)= {C_1(\mu)} \chi_{q,s}^A e^{i\phi_{q,s}^A}.
 \ed
The dimensionless parameters $\chi_{q,s}^{E,A}$ and
$\phi_{q,s}^{E,A}$ describe the strengths and the strong phases of
the involved matrix elements, where the superscripts $q,s$
differentiate the light quarks and the strange quarks strongly
produced in pairs. The $b$'s are defined for the $\pi\pi$ final
states, to which those for other final states are related via the
ratio of the decay constants $f_{P_1}f_{P_2}/f_\pi^2$. That is, the
flavor $SU(3)$ breaking effects from the strongly produced quark
pairs and from the decay constants have been taken into account. The
parameters $\chi_{q,s}^{E,A}$ and $\phi_{q,s}^{E,A}$ are then
assumed to be universal, and will be determined from data.
Similarly, we consider the $SU(3)$ breaking effect in the energy
release, which defines the scale of the Wilson coefficients
$C_{1,2}(\mu)$
 \be
 \mu=\sqrt{\Lambda m_D (1-r_1^2)(1-r_2^2)},
 \ed
with the mass ratios $r_{1,2}=m_{P_1,P_2}/m_D$.

As pointed out in \cite{Li:2009wba}, a special type of soft gluons,
called the Glauber gluons, may introduce an additional strong phase
to a nonfactorizable amplitude in two-body heavy-meson decays. The
Glauber phase associated with a pion is more significant, because of
its simultaneous role as a Nambu-Goldstone boson and a $q\bar q$
bound state: the valence quark and anti-quark of the pion are
separated by a short distance in order to reduce the confinement
potential energy, while the multi-parton states of the pion spread
over a huge space-time in order to meet the role of a massless
Nambu-Goldstone boson \cite{NS08}. The relatively larger soft effect
from the multi-parton states in the pion results in the Glauber
phase. Hence, we assign a phase factor $\exp(iS_{\pi})$ for each
pion involved in the nonfactorizable amplitudes $E$ and $A$. The
phase factor associated with the $\pi\pi$ final states is then given
by $\exp(2iS_{\pi})$ as explicitly shown in Appendix~\ref{app:amp}.
The reason why we do not introduce the Glauber phase into the
nonfactorizable emission diagrams, but only into the $W$-exchange
and $W$-annihilation diagrams is that the emission topology mainly
receives factorizable contributions in most of the decay modes.
Since the Glauber phase appears
only in nonfactorizable amplitudes, we will not include it into the
emission topology in the present work.

In summary, our parametrization for the emission amplitudes in
Eq.~(\ref{emit}) and the $W$-exchange and $W$-annihilation ones in
Eq.~(\ref{ann}) contains 12 free parameters. The global fit to all
the data of 28 $D\to PP$ branching ratios leads to the outcomes
 \be
&&\Lambda=0.56\;\;{\rm
GeV},~~\chi_{nf}=-0.59,~~\chi_q^E=0.11,~~\chi_s^E=0.18,~~\chi_q^A=0.12,~~\chi_s^A=0.17,
\non\\&&~~S_\pi=-0.50,~~\phi=-0.62,~~\phi_q^E=4.80
,~~\phi_s^E=4.23,~~\phi_q^A=4.06,~~\phi_s^A=3.48,\label{out}
 \ed
with the fitted $\chi^2=6.9$ per degree of freedom, which marks
a great improvement compared to $\chi^2=87$ in \cite{diag}. The parameter
$\Lambda=0.56$ GeV indeed corresponds to the soft momentum involved
in the $D$ meson, implying the typical energy release $\mu\sim 1$
GeV. This scale, which is not very low, indicates that the PQCD
factorization formulas \cite{LMS05} might provide a useful guideline
for formulating various contributions to the $D\to PP$ decays in our
framework. Since the fermion flows of the amplitudes $E$ and $A$ can
be converted into each other through the Fierz transformation, the
relations $\chi^E_q\sim\chi^A_q$, $\chi^E_s\sim\chi^A_s$, and the
similar relations for the strong phases are expected. The
satisfactory fit also confirms our postulation that the helicity
suppression works well, and the nonfactorizable contributions
dominate the $W$-exchange and $W$-annihilation amplitudes. A large
$SU(3)$ breaking effect has been revealed by the inequality
$\chi_q^{E(A)}\neq\chi_s^{E(A)}$. The values in Eq.~(\ref{out})
support $E>A$ in \cite{Fusheng:2011tw,diag} according to
Eq.~(\ref{b}), consistent with the fact that the $D^0$ meson
lifetime is shorter than the $D^+$ meson lifetime. The Glauber phase
$S_\pi=-0.50$ is in agreement with that
extracted in \cite{Li:2009wba}, which resolves the puzzle from the
dramatically different data for the direct CP asymmetries $A_{\rm
CP}(B^0\to\pi^\mp K^\pm)$ and $A_{\rm CP}(B^\pm\to\pi^0 K^\pm)$.

\begin{table}
\caption {Branching ratios in units of percentages for
Cabibbo-favored $D\to PP$ decays. Our results are compared to those
from the analysis including FSI effects \cite{pietro}, the
topological-diagram approach \cite{diag}, the pole-dominant model
\cite{Fusheng:2011tw}, and the experimental data
\cite{dataofPP}.}\label{CFPP}
\begin{tabular}{cccccccc}\hline\hline
 Modes & ~~~Br(FSI)~~~ & Br(diagram) &  ~~~~Br(pole)~~~~  &~~~~
Br(exp)~~~~~& ~~~~Br(this
 work)~~~\\\hline
 $D^0\to \pi^0\overline{K}^0$ & 1.35 & 2.36$\pm $0.08  & $2.4\pm0.7$ & 2.38$\pm $0.09    & 2.41 \\
$D^0\to \pi^+K^-$       & 4.03 & 3.91$\pm $0.17  & $3.9\pm1.0$ & 3.891$\pm $0.077  & 3.70 \\
$D^0\to \overline{K}^0\eta$  & 0.80 & 0.98$\pm $0.05  & $0.8\pm0.2$ & 0.96$\pm $0.06    & 1.00 \\
$D^0\to \overline{K}^0\eta'$ & 1.51 & 1.91$\pm $0.09  & $1.9\pm0.3$ & 1.90$\pm $0.11    & 1.73 \\
$D^+\to \pi^+\overline{K}^0$ & 2.51 & 3.08$\pm $0.36  & $3.1\pm2.0$ & 3.074$\pm $0.096  & 3.22 \\
$D_S^+\to K^+\overline{K}^0$ & 4.79 & 2.97$\pm $0.32  & $3.0\pm0.9$ & 2.98$\pm $0.08    & 3.00 \\
$D_S^+\to \pi^+\eta$    & 1.33 & 1.82$\pm $0.32  & $1.9\pm0.5$ & 1.84$\pm $0.15    & 1.65 \\
$D_S^+\to \pi^+\eta'$   & 5.89 & 3.82$\pm $0.36  & $4.6\pm0.6$ & 3.95$\pm $0.34    & 3.44 \\
$D_S^+\to \pi^+\pi^0$& & 0&0&$<$0.06& 0\\
\hline\hline\end{tabular} \end{table}

\begin{table}
\caption {Same as Table~\ref{CFPP} except for singly
Cabibbo-suppressed $D\to PP$ decays in units of
$10^{-3}$.}\label{SCSPP}
\begin{tabular}{cccccccc}\hline\hline
 Modes & ~~~Br(FSI)~~~ & Br(diagram) &  ~~~~Br(pole)~~~~  &~~~~
Br(exp)~~~~~& ~~~~Br(this
 work)~~~\\\hline
$D^0\to \pi^+\pi^-$  & 1.59 & 2.24$\pm $0.10 & $2.2\pm0.5$ & 1.45$\pm $0.05 & 1.43  \\
$D^0\to K^+K^-$      & 4.56 & 1.92$\pm $0.08 & $3.0\pm0.8$ & 4.07$\pm $0.10 & 4.19  \\
$D^0\to K^0\overline{K}^0$& 0.93 & 0&               $0.3\pm0.1$ & 0.320$\pm $0.038 & 0.36  \\
$D^0\to \pi^0\pi^0$  & 1.16 & 1.35$\pm $0.05 & $0.8\pm0.2$ & 0.81$\pm $0.05
& 0.57  \\
$D^0\to \pi^0\eta$   & 0.58 & 0.75$\pm $0.02 & $1.1\pm0.3$ & 0.68$\pm $0.07 & 0.94  \\
$D^0\to \pi^0\eta'$  & 1.7 & 0.74$\pm $0.02  & $0.6\pm0.2$ & 0.91$\pm $0.13 & 0.65  \\
$D^0\to \eta\eta$    & 1.0 & 1.44$\pm $0.08  & $1.3\pm0.4$ & 1.67$\pm $0.18 & 1.48  \\
$D^0\to \eta\eta'$   & 2.2 & 1.19$\pm $0.07  & $1.1\pm0.1$ &1.05$\pm $0.26  & 1.54  \\
$D^+\to \pi^+\pi^0$  & 1.7 & 0.88$\pm $0.10  & $1.0\pm0.5$ & 1.18$\pm $0.07 & 0.89  \\
$D^+\to K^+\overline{K}^0$& 8.6 & 5.46$\pm $0.53  & $8.4\pm1.6$ & 6.12$\pm $0.22 & 5.95  \\
$D^+\to \pi^+\eta$   & 3.6 & 1.48$\pm $0.26  & $1.6\pm1.0$ & 3.54$\pm $0.21 & 3.39  \\
$D^+\to \pi^+\eta'$  & 7.9 & 3.70$\pm $0.37  & $5.5\pm0.8$ & 4.68$\pm $0.29 & 4.58  \\
$D_S^+\to \pi^0K^+$  & 1.6 & 0.86$\pm $0.09  & $0.5\pm0.2$ & 0.62$\pm $0.23 & 0.67  \\
$D_S^+\to \pi^+K^0$  & 4.3 & 2.73$\pm $0.26  & $2.8\pm0.6$ & 2.52$\pm $0.27 & 2.21  \\
$D_S^+\to K^+\eta$   & 2.7 & 0.78$\pm $0.09  & $0.8\pm0.5$ & 1.76$\pm $0.36 & 1.00 \\
$D_S^+\to K^+\eta'$  & 5.2 & 1.07$\pm $0.17  & $1.4\pm0.4$ & 1.8$\pm $0.5   & 1.92  \\
\hline\hline\end{tabular} \end{table}

\begin{table}
\caption {Same as Table~\ref{CFPP} except for doubly
Cabibbo-suppressed $D\to PP$ decays in units of $10^{-4}$.
}\label{DCSPP}
\begin{tabular}{cccccccc}\hline\hline
 Modes & Br(diagram) &  ~~~~Br(pole)~~~~  &~~~~
Br(exp)~~~~~& ~~~~Br(this
 work)~~~\\\hline
$D^0\to \pi^0K^0$ & 0.67$\pm $0.02  & $0.6\pm0.2$ &                 & 0.69\\
$D^0\to \pi^-K^+$ & 1.12$\pm $0.05  & $1.6\pm0.4$ & 1.48$\pm $0.07  & 1.67\\
$D^0\to K^0\eta$  & 0.28$\pm $0.02  & $0.22\pm0.05$&                & 0.29\\
$D^0\to K^0\eta'$ & 0.55$\pm $0.03  & $0.5\pm0.1$ &                 & 0.50\\
$D^+\to \pi^+K^0$ & 1.98$\pm $0.22  & $1.7\pm0.5$ &                 & 2.38\\
$D^+\to \pi^0K^+$ & 1.59$\pm $0.15  & $2.2\pm0.4$ & 1.72$\pm $0.19  & 1.97\\
$D^+\to K^+\eta$  & 0.98$\pm $0.04  & $1.2\pm0.2$ & 1.08$\pm$0.17\footnote{Data from \cite{won}}&0.66\\
$D^+\to K^+\eta'$ & 0.91$\pm $0.17  & $1.0\pm0.1$ & 1.76$\pm$0.22\footnote{Data from \cite{won}}&1.14\\
$D_S^+\to K^+K^0$ & 0.38$\pm $0.04  & $0.7\pm0.4$ &                 & 0.63\\
\hline\hline\end{tabular} \end{table}

Our results for the $D\to PP$ branching ratios are presented in
Tables~\ref{CFPP}, \ref{SCSPP}, and \ref{DCSPP} for the
Cabibbo-favored, singly Cabibbo-suppressed, and doubly
Cabibbo-suppressed decays, respectively. The experimental data, as
well as the predictions from other approaches, such as the
topological-diagram approach \cite{diag}, the factorization approach
combined with the pole-dominant model for the $W$-exchange and
$W$-annihilation contributions \cite{Fusheng:2011tw}, and an
analysis with FSI effects from nearby resonances \cite{pietro}, are
also listed for comparison. It is obvious that the predictions for
the singly Cabibbo-suppressed decays, including the
$D^0\to\pi^+\pi^-$, $K^+K^-$, and $K^0\overline K^0$ modes, are well
consistent with the data: the known large $SU(3)$ breaking effects
in these decays have been properly described in our parametrization.
The major sources to $\chi^2$ of our fit come from the
$D^0\to\pi^0\pi^0$ ($\Delta\chi^2=23$) and $D^+\to\pi^+\pi^0$
($\Delta\chi^2=17$) decays. These modes are exceptional in the sense
that they involve the colo-suppressed emission amplitudes $C$, which
are dominated by the nonfactorizable contributions. To reduce $\chi^2$,
more $SU(3)$ breaking effects, such as the Glauber phase in $C$
\cite{Li:2009wba}, need to be introduced.

The Wilson coefficients for the $D^0\to\pi^+\pi^-$ and $K^+K^-$ decays,
 \be
& a_1(\pi\pi)=1.09,~~~& a_2(\pi\pi)=0.81e^{i147.8^\circ},
 \non\\
& a_1(KK)=1.10,~~~& a_2(KK)=0.83e^{i148.2^\circ},\label{a12}
 \ed
similar to those obtained in \cite{Fusheng:2011tw,diag},
and
 \be
C_2(\pi\pi)=1.26,~~~C_2(KK)=1.27,\label{c12}
 \ed
for the $W$-exchange and $W$-annihilation amplitudes indicate minor
$SU(3)$ breaking effects from the scale running.
As pointed out in \cite{Cheng:2012wr}, to
account for the $D^0\to \pi^+\pi^-$ and $D^0\to K^+K^-$ branching
ratios, their $W$-exchange amplitudes must be different in both the
magnitudes and strong phases. In our parametrization, the
distinction in the magnitudes is achieved by differentiating
$\chi_q$ from $\chi_s$ and the decay constant $f_\pi$ from $f_K$,
while the distinction in the strong phases is achieved by the
Glauber phase $S_\pi$. Another significant improvement appears in
the prediction for the $D^+\to\pi^+\eta$ branching ratio as
indicated in Table~\ref{SCSPP}. The branching ratios obtained in
\cite{diag,Fusheng:2011tw} are small, because the emission and
annihilation amplitudes extracted from the Cabibbo-favored processes
lead to a destructive interference in this Cabibbo-suppressed mode.
In our framework the additional Glauber phase associated with the
pion changes the destructive interference into a constructive one.
This provides another support for the $SU(3)$ breaking effect caused
by the Glauber phase.

The scale-dependent Wilson coefficients do improve the overall
agreement between the theoretical predictions and the
data involving the $\eta'$ meson. Taking the
$D^+\to\pi^+\eta'$ decay as an example, the corresponding parameters
are given by $a_1(\pi\eta')=1.12$ and
$a_2(\pi\eta')=0.89e^{i149.6^\circ}$ for the emission amplitudes
with the $\eta'$ meson emitted from the weak vertex, and
$C_2(\pi\eta')=1.32$ for the annihilation amplitudes. Compared to
Eqs.~(\ref{a12}) and (\ref{c12}), it is clear that the lower energy
release involved in the $D^+\to\pi^+\eta'$ decay with heavier final
states increases the Wilson coefficients. Benefited from the
scale-dependent Wilson coefficients, $\chi^2$ from all the $\eta'$
involved modes is reduced by nearly 10, relative to $\chi^2$ in the
previous fits in the pole model \cite{Fusheng:2011tw}.

\section{Direct CP Asymmetries}

In this section we predict the direct CP asymmetries in the $D\to
PP$ decays by combining the short-distance dynamics associated with
the penguin operators and the long-distance parameters in Eq.~(\ref{out}). The
direct CP asymmetry, defined by
 \be
 A_{\rm CP}(f)=\frac{\Gamma(D\to f)-\Gamma(\overline D\to\overline f)}
 {\Gamma(D\to f)+\Gamma(\overline D\to\overline
 f)},
 \ed
exists only in singly Cabibbo-suppressed modes.

\subsection{Tree-level CP violation}

A $D\to PP$ mode, receiving both contributions proportional to
$\lambda_d$ and $\lambda_s$, has the decay amplitude
 \be
 \mathcal A=\lambda_d \mathcal A_d+\lambda_s\mathcal A_s.
 \ed
The interference between the two terms in $\mathcal A$ leads to the
tree-level direct CP asymmetry
 \be
 A_{\rm CP}\approx-2\frac{{\rm Im}(\lambda_d^*\lambda_s)}
 {|\lambda_d|^2}\frac{{\rm Im}(\mathcal A_d^*\mathcal
 A_s)} {|\mathcal A_d-\mathcal A_s|^2},
 \ed
where the relation $\lambda_d+\lambda_s\approx0$ has been inserted
into the denominator. Except for the final states with the
$\eta^{(\prime)}$ meson, that contains both the $q\bar q$ and $s\bar
s$ components, four other channels $D^0\to K^0\overline K^0$,
$D^+\to K^+\overline K^0$, $D_s^+\to\pi^+K^0$, and $D_s^+\to\pi^0
K^+$ also exhibit the tree-level CP violation, though their values
have been found to be small \cite{Cheng:2012wr}. We can predict the
tree-level CP violation in the $D^0\to K^0\overline K^0$ decay as
indicated in the Table~\ref{CPV}, which is not attainable in the
diagrammatic approach in the $SU(3)$ limit \cite{Cheng:2012wr}. This
is one of the advantages of our formalism that has taken into account
most of $SU(3)$ breaking effects. The CP asymmetries in this mode,
including the CP violation in the $K^0$-$\overline K^0$ mixing,
would be possibly measured.

\subsection{Penguin-induced CP violation}

We extend our formalism to the penguin contributions to the $D\to
PP$ decays. For this purpose, the effective Hamiltonian for singly
Cabibbo-suppressed charm decays
 \be
 \mathcal H_{\rm eff}={G_F\over \sqrt 2}
 \left[\sum_{q=d,s}V_{cq}^*V_{uq}(C_1(\mu)O_1^q(\mu)+C_2(\mu)O_2^q(\mu))
 -V_{cb}^*V_{ub}\left(\sum_{i=3}^6C_i(\mu)O_i(\mu)+C_{8g}(\mu)O_{8g}(\mu)\right)\right],
 \ed
is considered, with the QCD penguin operators
 \be
 O_3&=&\sum_{q'=u,d,s}(\bar u_\alpha c_\alpha)_{V-A}(\bar q'_\beta
 q'_\beta)_{V-A},~~~
 O_4=\sum_{q'=u,d,s}(\bar u_\alpha c_\beta)_{V-A}(\bar q'_\beta q'_\alpha)_{V-A},
 \non\\
 O_5&=&\sum_{q'=u,d,s}(\bar u_\alpha c_\alpha)_{V-A}(\bar q'_\beta
 q'_\beta)_{V+A},~~~
 O_6=\sum_{q'=u,d,s}(\bar u_\alpha c_\beta)_{V-A}(\bar q'_\beta
 q'_\alpha)_{V+A},
 \ed
and the chromomagnetic penguin operator,
\begin{eqnarray}
O_{8g}=\frac{g}{8\pi^2}m_c{\bar
u}\sigma_{\mu\nu}(1+\gamma_5)T^aG^{a\mu\nu}c,
\end{eqnarray}
$T^a$ being a color matrix. The explicit expressions for the Wilson
coefficients $C_{3-6}(\mu)$ in terms of the running coupling
constant $\alpha_s(\mu)$ are given in Appendix~\ref{app:wc}.

\begin{figure}
\begin{center}
\includegraphics[scale=0.5]{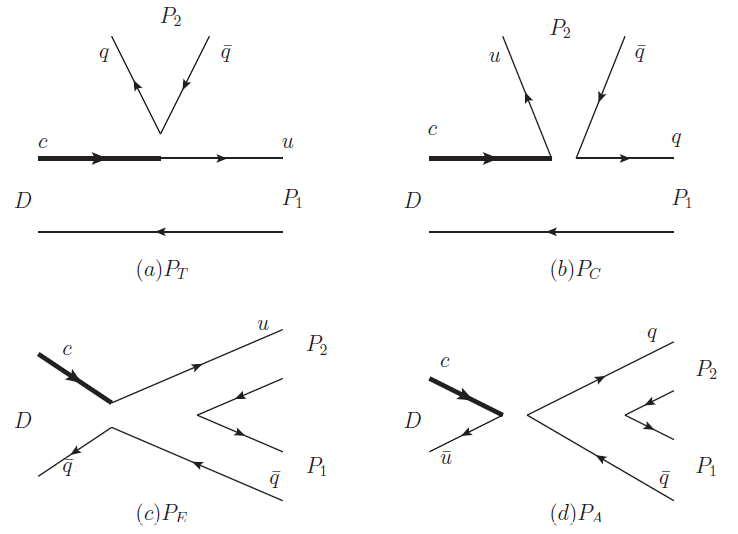}
\end{center}
\caption{Four topological penguin diagrams contributing to $D\to PP$
decays: (a) the penguin color-favored emission amplitude $P_T$, (b)
the penguin color-suppressed emission amplitude $P_C$, (c) the
penguin exchange amplitude $P_E$, and (d) the penguin annihilation
amplitude $P_A$. Note that $P_T(P_C)$ is not truly color-favored
(color-suppressed), but possesses the same topology as $T(C)$.
}\label{fig:topopen}
\end{figure}

Below we sort out the short-distance dynamics associated with the
above penguin operators. The contributions from the $(V-A)(V-A)$
current in the operators $O_{3,4}$ can be directly related to those
from the tree operators by replacing the Wilson coefficients. The
relations between the hadronic matrix elements from the $(V-A)(V+A)$
or $(S+P)(S-P)$ current and from the $(V-A)(V-A)$ current are
subtler. For the penguin color-favored emission amplitudes, we have
 \be
 P_T&=&a_3(\mu)\la P_2|(\bar q
 q)_{V-A}|0\ra\la P_1|(\bar u c)_{V-A}|D\ra
 +a_5(\mu)\la P_2|(\bar q
 q)_{V+A}|0\ra\la P_1|(\bar u c)_{V-A}|D\ra,
 \non\\
 &=&\left[a_3(\mu)-a_5(\mu)\right]
 f_{P_2}(m_D^2-m_{P_1}^2)F_0^{DP_1}(m_{P_2}^2),
 \ed
with the Wilson coefficients
\begin{eqnarray}
& &a_3(\mu)=C_3(\mu)+{C_4(\mu)\over N_c},\nonumber\\
& &a_5(\mu)=C_5(\mu)+{C_6(\mu)\over N_c},
\end{eqnarray}
where the minus sign in front of the Wilson coefficient $a_5$ arises
from the equality $\la P_2|(\bar q q)_{V+A}|0\ra=-\la P_2|(\bar q
q)_{V-A}|0\ra$.

We write the penguin color-suppressed emission amplitude
as
  \be\label{pc}
 P_C&=&a_4(\mu)\la P_2|(\bar
 u q)_{V-A}|0\ra\la P_1|(\bar q c)_{V-A}|D\ra
 -2a_6(\mu)\la P_2|(\bar
 u q)_{S+P}|0\ra\la P_1|(\bar q c)_{S-P}|D\ra
 \non\\
&=&\left[a_4(\mu)+a_6(\mu)r_\chi\right]
f_{P_2}(m_D^2-m_{P_1}^2)F_0^{DP_1}(m_{P_2}^2),
 \ed
with the Wilson coefficients
\begin{eqnarray}
& &a_4(\mu)=C_4(\mu)+C_3(\mu)\left[{1\over
N_c}+\chi_{nf}e^{i\phi}\right],\nonumber\\
& &a_6(\mu)=C_6(\mu)+{C_5(\mu)}\left[{1\over
N_c}+\chi_{nf}e^{i\phi}\right].\label{a46}
\end{eqnarray}
The chiral factor
 \be\label{chiral}
 r_\chi={2m_{P_2}^2\over m_c(m_u+m_q)},
 \ed
with the $u$-quark ($q$-quark) mass $m_u$ ($m_q$), exhibits formal
suppression by a power of $1/m_c$, but takes a value of order
unity actually. Note that the $(S-P)(S+P)$ nonfactorizable amplitude cannot
be simply related to the $(V-A)(V-A)$ nonfactorizable amplitude.
However, the PQCD factorization formulas for heavy-meson decays
\cite{LMS05} reveal strong similarity between them in the dominant small
parton momentum region, which supports the parametrization in
Eq.~(\ref{a46}). The chiral factor for the
$\eta^{(\prime)}$ meson receives an additional contribution from the
axial anomaly, as illustrated in Appendix~\ref{anomaly}.

The treatment of the penguin exchange amplitude $P_E$ and the
penguin annihilation amplitude $P_A$ demands a more careful
formulation. It is obvious that the matrix elements of the operator
$O_3$ ($O_4$) are similar to those of $O_2$ ($O_1$). As discussed in
Sec.~II, the factorizable contributions to the above matrix elements
vanish in the $SU(3)$ limit because of the helicity suppression, and
have been neglected. Considering only the nonfactorizable
contributions, $O_4$ gives rise to part of $P_A$. The
operator $O_6$ also contributes to $P_A$, to which the helicity
suppression applies. Therefore, its matrix elements can be combined
with those of $O_4$ under the equality
\begin{eqnarray}
\la P_1(q\bar q')P_2(q'\bar q)|(\bar uc)_{V-A}(\bar q'
 q')_{V+A}|D(c\bar u)\ra=\la P_1(q\bar q')P_2(q'\bar q)|(\bar uc)_{V-A}(\bar q'
 q')_{V-A}|D(c\bar u)\ra,\label{eqe}
\end{eqnarray}
which has been verified by the PQCD factorization formulas
\cite{LMS05}. Equation~(\ref{eqe}) is attributed to the fact that
only the vector piece $V$ in the $(\bar q' q')_{V\pm A}$ currents
contributes to the production of two pseudoscalar mesons. Thus the
QCD-penguin annihilation amplitude is parameterized as
 \be
 P_A=[C_4(\mu)+C_6(\mu)]\chi^A_{q,s}e^{i\phi^A_{q,s}}f_Dm_D^2\left({f_{P_1}f_{P_2}\over
 f_\pi^2}\right).
 \ed

For the amplitude $P_E$, the helicity suppression does not apply to
the matrix elements of $O_{5,6}$, so the factorizable contributions
exist. The nonfactorizable contribution from $O_5$ vanishes in the
$SU(3)$ limit, as verified by the PQCD factorization formulas
\cite{LMS05}, and will be neglected. We then apply the Fierz
transformation and the factorization hypothesis to
\begin{eqnarray}
\la P_1(q'\bar q)P_2(u\bar q')|(\bar uc)_{V-A}(\bar q'
 q')_{V+A}|D(c\bar q)\ra
=-2\la P_1(q'\bar q)P_2(u\bar q')|(\bar u q')_{S+P}|0\ra \la 0|(\bar
q'c)_{S-P}|D(c\bar q)\ra.
\end{eqnarray}
The scalar form factor involved in the matrix element
$\la P_1(q'\bar q)P_2(u\bar q')|(\bar u q')_{S+P}|0\ra$ has been assumed
to be dominated by scalar resonances, and fixed in \cite{Choi:1998yx}.
It has been demonstrated that the similar
formalism works for the analysis of the $\tau\to K\pi \nu_\tau$
decay \cite{Choi:1998yx}.
Combining the nonfactorizable contribution from $O_3$, $P_E$ is written as
 \be
 P_E=C_3(\mu)\chi^E_{q,s}e^{i\phi^E_{q,s}}f_Dm_D^2\left({f_{P_1}f_{P_2}\over
 f_\pi^2}\right)+2\left[C_6(\mu)+\frac{C_5(\mu)}{N_c}\right]
 g_SB_S(m_D^2)m_S\bar f_S f_D{m_D^2\over
 m_c},\label{pe}
 \ed
where $g_{S}$ is an effective strong coupling constant, $\bar f_S$
and $m_S$ are the decay constant and the mass of the scalar
resonance particle $S$, respectively, and the function $B_S$
represents the Breit-Wigner propagator of the scalar resonance,
 \be
 B_{S}(q^2)={1\over
 q^2-m_{S}^2+im_{S}\Gamma_{S}(q^2)}.\label{bs}
 \ed
In the above expression $q^2$ denotes the invariant mass squared of
the $P_1P_2$ final state, and the strong phase from the width $\Gamma_S(q^2)$
provides a major source to direct CP asymmetries.
More detail for the treatment of the
scalar matrix element $\la P_1(q'\bar q)P_2(u\bar q')|\bar u
q'|0\ra$ can be found in Appendix~\ref{me}.

\subsection{Quark Loops and Magnetic Penguin}

\begin{figure}[htpb]
\begin{center}
\includegraphics[scale=0.5]{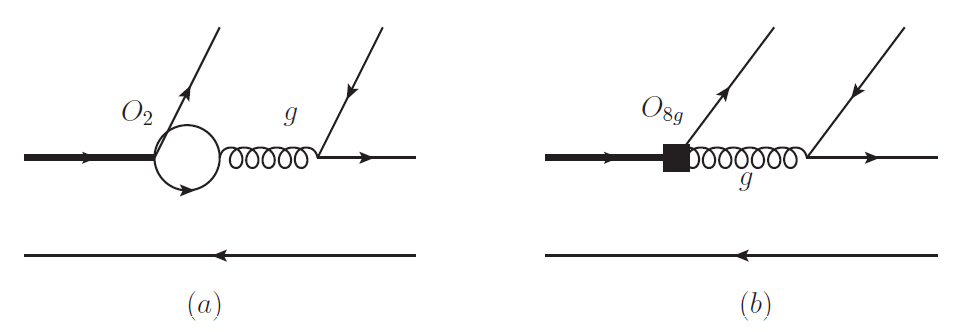}
\end{center}
\caption{Quark-loop and magnetic-penguin contributions
to $D\to PP$ decays.}\label{fig:qlmg}
\end{figure}

To complete the penguin contributions to the $D\to PP$ decays,
we include the quark loops from the tree operators and the
magnetic penguin as displayed Fig.~\ref{fig:qlmg}.
The former up to next-to-leading order in the
coupling constant can be absorbed into the Wilson coefficients
\cite{BBNS99}
\begin{eqnarray}
C_{3,5}(\mu) &\to& C_{3,5}(\mu) - \frac{\alpha_s(\mu)}{8\pi
N_c}\sum_{q=d,s}\frac{\lambda_{q}}
{\lambda_{b}}C^{(q)}(\mu,\langle l^2\rangle),\nonumber\\
C_{4,6}(\mu) &\to& C_{4,6}(\mu) +
\frac{\alpha_s(\mu)}{8\pi}\sum_{q=d,s}\frac{\lambda_{q}}
{\lambda_{b}}C^{(q)}(\mu,\langle l^2\rangle), \label{eq:CPfact}
\end{eqnarray}
with $\langle l^2\rangle$ being the averaged invariant mass
squared of the virtual gluon emitted from the quark loop. The
function $C^{(q)}$ is written as
\begin{eqnarray}
C^{(q)}(\mu,\langle l^2\rangle)= \left[ G^{(q)}(\mu,\langle l^2\rangle)  - \frac{2}{3}
\right] C_2(\mu),
\end{eqnarray}
with the function
\begin{eqnarray}\label{Gint}
G^{(q)}(\mu,\langle l^2\rangle) &=& - 4 \int_0^1 dx\, x (1-x)
\ln\frac{m_q^2 - x(1-x) \langle l^2\rangle}{\mu^2}.\label{gc}
\end{eqnarray}
Because of the smallness of
the associated CKM matrix elements, we have ignored the quark loops
from the penguin operators.

Using the unitarity relation of the CKM matrix, we reexpress the
second term in Eq.~(\ref{eq:CPfact}) as
 \be
\frac{\alpha_s(\mu)}{8\pi}\frac{\lambda_d C^{(d)}+\lambda_s C^{(s)}}
{\lambda_b}=\frac{\alpha_s(\mu)}{8\pi}\left[{\lambda_d\over\lambda_b}
\left(C^{(d)}-C^{(s)}\right)-C^{(s)}\right].\label{lambda}
 \ed
The term proportional to $\lambda_d$ contributes to the tree
amplitudes, which is, however, numerically negligible,
$(\alpha_s/8)\pi(C^{(d)}-C^{(s)})\sim O(10^{-4})\ll C_{1,2}$.
It is the reason why we did not include
the quark loops in the analysis of the
branching ratios. Keeping only the second term in Eq.~(\ref{lambda}),
Eq.~(\ref{eq:CPfact}) reduces to
\begin{eqnarray}
C_{3,5}(\mu) &\to& C_{3,5}(\mu) + \frac{\alpha_s(\mu)}{8\pi
N_c}C^{(s)}(\mu,\langle l^2\rangle),\nonumber\\
C_{4,6}(\mu) &\to& C_{4,6}(\mu) -
\frac{\alpha_s(\mu)}{8\pi}C^{(s)}(\mu,\langle
l^2\rangle).\label{quark}
\end{eqnarray}
We have confirmed that no much numerical difference arises from
replacing $C^{(s)}$ by $C^{(d)}$.

The magnetic-penguin contribution can be included into the Wilson
coefficients for the penguin operators following the substitutions
\cite{BBNS99}
\begin{eqnarray}
C_{3,5}(\mu)&\to& C_{3,5}(\mu) + \frac{\alpha_s(\mu)}{8\pi N_c}
\frac{2m_c^2}{\langle l^2\rangle}C_{8g}^{\rm eff}(\mu),\nonumber\\
C_{4,6}(\mu)&\to& C_{4,6}(\mu) - \frac{\alpha_s(\mu)}{8\pi }
\frac{2m_c^2}{\langle l^2\rangle}C_{8g}^{\rm eff}(\mu),\label{mag}
\end{eqnarray}
with the effective Wilson coefficient $C_{8g}^{\rm eff}=C_{8g}+C_5$.
To predict the direct CP asymmetries, we adopt
$\langle l^2\rangle\approx (P_1/2+P_2/2)^2 \approx m_D^2/4$. This
choice of $\langle l^2\rangle$ corresponds to the kinematic
configuration, where the spectator in the decay takes half
of the $P_1$ meson momentum, consistent with the scale
parametrization in Eq.~(\ref{scale}), and a valence quark in the
$P_2$ meson also carries half of its momentum. We have varied
$\langle l^2\rangle$ between $m_D^2/2$ and $m_D^2/10$, and
observed only slight change in our predictions for the direct CP
asymmetries. The
contribution from the magnetic penguin is much smaller than that from
the quark loops, $C_{8g}^{\rm eff}\ll C^{(q)}(\mu,\la l^2\ra)$.
It turns out that the integral in Eq.~(\ref{Gint}) generates a large
imaginary part as $m_q^2\ll \la l^2\ra$, and a
sizable shift of the strong phases in the penguin amplitudes.

It is seen that all the important penguin amplitudes have been
formulated without introducing additional free parameters.
Compared with \cite{Brod:2011re}, the similarity
is that the effective Wilson coefficients in Eqs.~(\ref{quark}) and
(\ref{mag}) have been employed. The difference appears in the
derivation of the involved hadronic matrix elements. Both the
leading-power and subleading-power penguin contributions,
including their strengths and strong phases, are determined
in our formalism. The leading-power penguin contributions were
calculated in QCDF, but the subleading-power ones were estimated
in the large $N_c$ assumption in \cite{Brod:2011re}.
This is the reason why only the strengths of the penguin contributions
were obtained in \cite{Brod:2011re}.

\subsection{Numerical Results}

We adopt $m_u=m_d=\overline m_q=5.3$ MeV, and $m_s=136$ MeV for the quark
masses at the scale 1 GeV in the numerical analysis. The inputs for
the involved decay constants and transition form factors are taken
to be the same as in \cite{Fusheng:2011tw}. In most modes
$P_C$ which is enhanced by the chiral factor, and $P_E$ from the
$(S-P)(S+P)$ current
are the major QCD-penguin amplitudes. Our predictions for the
direct CP asymmetries in the $D\to PP$ decays are listed in
Table~\ref{CPV}, which differ from those derived in other
approaches, and can be confronted with future data.  The result
for $D^0\to K^0\overline K^0$ is dominated by the tree-level CP
violation, due to smallness of the involved penguin annihilation
contribution. If considering the isospin symmetry breaking from
unequal $u$ and $d$ quark masses, the chiral factors for the
$u\bar u$ and $d\bar d$ components of the $\pi^0$ meson will be
different. The corresponding penguin contributions then do not cancel
exactly, and a nonzero direct CP asymmetry would appear in the
$D^+\to\pi^+\pi^0$ decay. If turning off the Glauber phase in the
$D^0\to\pi^+\pi^-$ decay, the direct CP asymmetry $A_{\rm CP}^{\rm
tot}(\pi^+\pi^-)$ decreases from $0.58\times 10^{-3}$ to $0.49\times
10^{-3}$. The 30\% change in the branching ratio (see
Table~\ref{SCSPP}) and the 15\% change in the direct CP asymmetry
indicate that the Glauber phase is not a negligible mechanism.

We predict the difference between the direct CP asymmetries in the
$D^0\to K^+K^-$ and $D^0\to\pi^+\pi^-$ decays,
\begin{eqnarray}
\Delta A_{\rm CP}=-1.00\times10^{-3},\label{pre}
\end{eqnarray}
which is in the same sign as the data, but an order of magnitude
smaller than the central value of the LHCb data in Eq.~(\ref{lhcb}),
and lower than the CDF data in Eq.~(\ref{cdfn}).
To check whether our prediction makes sense, we write $\Delta A_{\rm CP}$ as
 \be
 \Delta A_{\rm CP} =-2 r \sin\gamma
 \left({|\mathcal P^{KK}|\over|\mathcal T^{KK}|}\sin\delta^{KK}+
 {|\mathcal P^{\pi\pi}|\over|\mathcal
 T^{\pi\pi}|}\sin\delta^{\pi\pi}\right),
 \ed
with $r=|\lambda_b|/|\lambda_d|=|\lambda_b|/|\lambda_s|$,
$\lambda_q=V_{cq}^*V_{uq}$, $\gamma$ being the CP violating weak
phase, and $\delta^{KK}$ ($\delta^{\pi\pi}$) being the relative
strong phase between the tree amplitude $\mathcal{T}^{KK}$
($\mathcal{T}^{\pi\pi}$) and the penguin amplitude
$\mathcal{P}^{KK}$ ($\mathcal{P}^{\pi\pi}$) for the $D^0\to K^+K^-$
($D^0\to \pi^+\pi^-$) mode. It is easy to find
$r\approx0.7\times10^{-3}$ and $\gamma=(73^{+22}_{-25})^\circ$ from
the Particle Data Group 2010 \cite{PDG}, i.e.,
$2r\sin\gamma\approx1.3\times10^{-3}$. For $|\mathcal P|/|\mathcal
T|\sim O(1)$ and $\sin\delta\sim1$, $\Delta A_{\rm CP}$ could reach
only a few times of $10^{-3}$. The topological amplitudes for the
$D^0\to \pi^+\pi^-$ and $K^+K^-$ modes obtained in this work are
given, in unit of $10^{-6}$ GeV, by
 \be
 T^{\pi\pi}&=&2.73,~~~E^{\pi\pi}=0.82 e^{-i 142^{\circ}},
 \non\\
 T^{KK}&=&3.65,~~~E^{KK}=1.2 e^{-i 85^\circ},
 \ed
  \be
 P_C^{\pi\pi}=0.87 e^{i 134^\circ},~~~~P_E^{\pi\pi}&=&0.81 e^{i 111^\circ},~~~~~P_A^{\pi\pi}=0.25 e^{-i
 43^{\circ}},
 \non\\
 P_C^{KK}=1.21 e^{i 135^\circ},~~~P_E^{KK}&=&0.87 e^{i 111^\circ},~~~~P_A^{KK}=0.45
 e^{-i5^\circ},
 \ed
which lead to the total tree and penguin amplitudes and their ratios
 \be
 \mathcal{T}^{\pi\pi}=2.14 e^{-i
 14^\circ},~~~\mathcal{P}^{\pi\pi}&=&1.40 e^{i
 121^\circ},~~~{\mathcal{P}^{\pi\pi}\over\mathcal{T}^{\pi\pi}} =0.66 e^{i134^\circ},
 \non\\
 \mathcal{T}^{KK}=3.94 e^{-i
 18^\circ},~~~\mathcal{P}^{KK}&=&1.79 e^{i
 114^\circ},~~~{\mathcal{P}^{KK}\over\mathcal{T}^{KK}} =0.45
 e^{i131^\circ}.
 \ed

\begin{table}\caption
{Direct CP asymmetries in $D\to PP$ decays in the unit of
($10^{-3}$).  $A^{\rm tree}_{\rm CP}$ denotes the tree-level CP
asymmetry and $A_{\rm CP}^{\rm tot}$ denotes the CP asymmetry
arising from the interference between the total tree and penguin
amplitudes. Results from the analysis including FSI effects
\cite{pietro}, and from the topological-diagram approach
\cite{Cheng:2012wr} are presented for comparison.}\label{CPV}
\begin{tabular}{cccccccc}\hline\hline
 Modes & $A_{\rm CP}$(FSI) & $A_{\rm CP}$(diagram) &  ~~~$A_{\rm CP}^{\rm tree}$~~~ &
~~~$A_{\rm CP}^{\rm tot}$~~~\\\hline
$D^0\to \pi^+\pi^-$  & 0.02$\pm$0.01  &   0.86  & 0     & 0.58   \\
$D^0\to K^+K^-$      & 0.13$\pm$0.8   & $-$0.48 & 0     & $-0.42$  \\
$D^0\to \pi^0\pi^0$  & $-0.54\pm$0.31 &   0.85  & 0     & 0.05  \\
$D^0\to K^0\overline{K}^0$& $-0.28\pm0.16$ &0   & $1.11$  & $1.38$  \\
$D^0\to \pi^0\eta$   & 1.43$\pm$0.83  & $-$0.16 &$-0.33$& $-0.29$  \\
$D^0\to \pi^0\eta'$  & $-0.98\pm$0.47 & $-$0.01 & $$0.53  & 1.53  \\
$D^0\to \eta\eta$    & 0.50$\pm$0.29  & $-$0.71 & $$0.29  & $0.18$  \\
$D^0\to \eta\eta'$   & $0.28\pm0.16$  &   0.25  &$-0.30$& $-0.94$  \\
$D^+\to \pi^+\pi^0$  &                &   0     &     0 & 0  \\
$D^+\to K^+\overline K^0$&$-0.51\pm$0.30&$-$0.38&$-0.13$& $-$0.93  \\
$D^+\to \pi^+\eta$   &                & $-$0.65 &$-0.54$& $-0.26$  \\
$D^+\to \pi^+\eta'$  &                &   0.41  & $$0.38  & $1.18$  \\
$D_S^+\to \pi^0K^+$  &                &   0.88  & $$0.32  & $$0.39  \\
$D_S^+\to \pi^+K^0$  &                &   0.52  &$0.13$& 0.84  \\
$D_S^+\to K^+\eta$   &                & $-$0.19 & $$0.80  &$$ 0.70 \\
$D_S^+\to K^+\eta'$  &                & $-$0.41 &$-0.45$ & $-1.60$  \\
\hline\hline\end{tabular} \end{table}

It is a concern for our formalism whether all the important nonperturbative
sources in $D$ meson decays have been identified. If not, the simple
replacement of the Wilson coefficients may not work for estimating
the penguin contributions. As explained in the previous section, the
outcomes in Eq.~(\ref{out}) are all reasonable, so most of the
important nonperturbative sources should have been taken into
account. The uncertainty of our formalism arises mainly from the
parametrization of the scalar form factors in $P_E$. The difference
of the direct CP asymmetries changes from $-1.0\times10^{-3}$ to
$-1.6\times10^{-3}$, when the effective strong coupling
$g_S=g_{K^*K\pi}$ is increased by a factor of 2.
If varying the strong phases in the penguin amplitudes
arbitrarily, $\Delta A_{\rm CP}$ would be at most
$-1.4\times10^{-3}$. Varying the input parameters, such as the
quark masses $\overline m_q=4.0\sim6.5$ MeV and $m_s= 108\sim 176$ MeV,
the mass $M_{f_0(1370)}= 1200\sim 1500$ MeV and the width
$\Gamma_{f_0(1370)}=200\sim500$ MeV of the dominant scalar resonance $f_0(1370)$,
and the CKM matrix elements $|V_{ub}|=(3.89\pm0.44)\times10^{-3}$ and
$\gamma=(73^{+22}_{-25})^\circ$, we have $\Delta
A_{CP}=(-0.57\sim-1.87)\times10^{-3}$. In conclusion, if the central values
of the LHCb and CDF data persist, they can be regarded as a signal
of new physics. At this moment, there is plenty of room for
accommodating the LHCb and CDF data by means of models extended
beyond the SM. Those models with additional sources of CP violation
are all suitable candidates, like supersymmetry models, left-right
models, models with the leptoquark, the fourth generation, or the
diquark, and so on. New physics effects on the direct CP asymmetries
in $D\to PP$ decays can be explored by combining corresponding
short-distance Wilson coefficients with the long-distance parameters
determined in this work.

\section{Conclusion}

In this paper we have proposed a theoretical framework for analyzing
two-body nonleptonic $D$ meson decays, based on the factorization of
short-distance (long-distance) dynamics into Wilson coefficients
(hadronic matrix elements of four-fermion operators). Because of the
small charm quark mass just above 1 GeV, a perturbative theory for
the hadronic matrix elements may not be valid. Our idea is to identify
as complete as possible the important sources of nonperturbative dynamics
in the decay amplitudes, and parameterize them in the framework of the
factorization hypothesis: we have considered the evolution of the Wilson
coefficients with the energy release in individual decay modes, which
depends on the scale $\Lambda$ of the soft degrees of freedom in the $D$
meson. The hadronic matrix elements of the four-fermion operators have
been parameterized into the strengths $\chi$'s and the strong phases
$\phi$'s. It is crucial to introduce the Glauber strong phase $S_\pi$
associated with a pion in the nonfactorizable annihilation amplitudes,
which is due to the unique role of the pion as a Nambu-Goldstone boson and a
quark-anti-quark bound state simultaneously. The flavor $SU(3)$ symmetry
breaking effects have been taken into account in the above framework
appropriately. Fitting our parametrization
to the abundant data of the $D\to PP$ branching ratios, all the
nonperturbative parameters have been determined.

It has been shown that our framework greatly improves the global fit to the
measured $D\to PP$ branching ratios: the evolution of the Wilson
coefficients improves the overall agreement with the data involving
the $\eta'$ meson, and the Glauber phase resolves the long-standing
puzzle from the $D^0\to\pi^+\pi^-$ and $D^0\to K^+K^-$ branching
ratios. The Glauber phase has also enhanced the predicted $D^+\to\pi^+\eta$
branching ratio, giving much better consistency with the data.
The value of $\Lambda$ is in the correct order of
magnitude for characterizing the soft degrees of freedom in the $D$
meson, and the Glauber phase $S_\pi$ agrees with that extracted from
the data for the direct CP asymmetries in the $B\to\pi K$ decays.
The similarity of the $\chi$ values for the $W$-exchange and for the
$W$-annihilation conforms our expectation, because
these two topologies can be converted to each other via the Fierz
transformation. The difference between the $\chi$ values for the
strongly produced light-quark and strange-quark pairs implies the
significant $SU(3)$ breaking effects in the $D\to PP$ decays.

Once having determined the nonperturbative parameters from the fit
to the branching ratios, the replacement of the Wilson
coefficients works for estimating the penguin contributions. For
those penguin amplitudes, which cannot be related to tree amplitudes
through the above replacement, we have shown that they are either
factorizable or suppressed by the helicity conservation. If they are factorizable,
such as the scalar penguin annihilation contribution, data from
other processes can be used for their determination. We are
then able to predict the direct CP asymmetries in $D\to PP$ decays
without ambiguity. It has been found that the strong phases from
the quark loops and the scalar penguin annihilation dominate the
direct CP asymmetries. Many of our predictions in
Table~\ref{CPV} can be confronted with future data. In particular,
we have predicted $\Delta A_{\rm CP}=-1.00\times 10^{-3}$, which is
lower than the LHCb and CDF data. As pointed out at the end of the
previous section, the uncertainty of our formalism arises mainly
from the parametrization of the scalar form factors in $P_E$.
However, even increasing the associated effective strong coupling
by a factor or 2, $\Delta A_{\rm CP}$ remains of order $10^{-3}$ .
We conclude that the LHCb and CDF data will reveal a new-physics
signal, if their central values persist.

\section*{Acknowledgment}
We are grateful to Hai-Yang Cheng, Emi Kou, Satoshi Mishima,
Nejc Kosnik, Svjetlana Fajfer, and Andrey Tayduganov
for useful discussions. This work was supported in part by
the National Science Council of R.O.C. under Grant No.
NSC-98-2112-M-001-015-MY3, by the National Center
for Theoretical Sciences, and by National Science Foundation
of China under the Grant No. 11075168.

\begin{appendix}
\section{Amplitudes}\label{app:amp}

Different decay modes receive contributions from different
topological amplitudes. Take the $D^0\to\pi^+\pi^-$ and $D^0\to
K^+K^-$ modes as examples, whose expressions are give by
 \be\label{eq:amp}
 \mathcal A(D^0\to\pi^+\pi^-)&=&{G_F\over\sqrt 2}
 \bigg[\lambda_d\left(T+E\right)-\lambda_b
 \left(P_C+2P_A^d+P_E^d\right)\bigg],
 \non\\&=&{G_F\over\sqrt 2}\bigg\{V_{cd}^*V_{ud}
 \left[a_1(\mu)(m_D^2-m_\pi^2)f_\pi F^{D\pi}_0(m_\pi^2)
 +C_2(\mu)e^{i(\phi_q^E+2S_\pi)}\chi_q^E
 f_Dm_D^2\right]
 \non\\
 &&-V_{cb}^*V_{ub}\bigg[a_{P_C}(\mu)(m_D^2-m_\pi^2)
 f_\pi F_0^{D\pi}(m_\pi^2) +2
 a_{P_A}(\mu)e^{i(\phi_q^A+2S_\pi)}\chi_q^Af_Dm_D^2
 \non\\
 &&+C_3(\mu)e^{i(\phi_q^E+2S_\pi)}\chi_q^Ef_Dm_D^2+2a'_6(\mu)
 g_S f_D{m_D^2\over m_c}\sum_{f_0}B_{f_0}(m_D^2)m_{f_0}
 \bar f_{f_0}\bigg]\bigg\},
 \non\\
  \mathcal A(D^0\to K^+K^-)&=&{G_F\over\sqrt 2}\bigg[\lambda_s
  \left(T+E\right)-\lambda_b\left(P_C+P_A^u+P_A^s+P_E^s\right)\bigg],
  \non\\
  &=&{G_F\over\sqrt 2}\bigg\{V_{cs}^*V_{us}
 \left[a_1(\mu)(m_D^2-m_K^2)f_K F^{DK}_0(m_K^2)+C_2(\mu)e^{i\phi_q^E}
 \chi_q^E f_Dm_D^2{f^2_K\over f^2_\pi}\right]
 \non\\
 &&-V_{cb}^*V_{ub}\bigg[a_{P_C}(\mu)(m_D^2-m_K^2)f_K F_0^{DK}(m_K^2) +
 a_{P_A}(\mu)e^{i\phi_q^A}\chi_q^Af_Dm_D^2{f^2_K\over f^2_\pi}
 \non\\
 &&+ a_{P_A}(\mu)e^{i\phi_s^A}\chi_s^Af_Dm_D^2{f_K^2\over f_\pi^2}
 +C_3(\mu)e^{i\phi_s^E}\chi_s^Ef_Dm_D^2{f^2_K\over f^2_\pi}
 \non\\
 &&+2a'_6(\mu)g_S f_D{m_D^2\over m_c}\sum_{f_0}B_{f_0}(m_D^2)m_{f_0}
 \bar f_{f_0}\bigg]\bigg\},
 \ed
with the coefficients
  \be
 & &a_{P_C}(\mu)=\left[a_4(\mu)
 +a_6(\mu)r_\chi\right],
 \non\\
 & &a_{P_A}(\mu)=C_4(\mu)+C_6(\mu),\non\\
 & &a'_6(\mu)=C_6(\mu)+\frac{C_5(\mu)}{N_c}.
 \ed
Notice the Glauber phase factors $e^{i2S_\pi}$ associated with
the nonfactorizable amplitudes $\chi_q^E$ and $\chi_q^A$ in Eq.~(\ref{eq:amp}).

\section{Wilson coefficients}\label{app:wc}

In this Appendix we present the evolution of the Wilson coefficients
in the scale $\mu<m_c$ \cite{Fajfer:2002gp},
 \be
 C_1(\mu)&=&0.2334\alpha^{1.444}+0.0459\alpha^{0.7778}
 -1.313\alpha^{0.4444}+0.3041\alpha^{-0.2222},
 \\
 C_2(\mu)&=&-0.2334\alpha^{1.444}+0.0459\alpha^{0.7778}
 +1.313\alpha^{0.4444}+0.3041\alpha^{-0.2222},
 \\
 C_3(\mu)&=&0.0496\alpha^{-0.2977}-0.0608\alpha^{-0.2222}+0.0025\alpha^{-0.1196}
 -1.23\alpha^{0.4173}
 \non\\&&+1.313\alpha^{0.4444}-0.0184\alpha^{0.7023}-0.0076\alpha^{0.7778}
 -0.0188\alpha^{0.8025}
 \non\\&&+0.0018\alpha^{0.8804}+0.2362\alpha^{1.417}
 -0.1986\alpha^{1.444}+0.0004\alpha^{1.802},
 \\
  C_4(\mu)&=&0.075\alpha^{-0.2977}-0.0608\alpha^{-0.2222}+0.0012\alpha^{-0.1196}
 +1.179\alpha^{0.4173}
 \non\\&&-1.313\alpha^{0.4444}+0.0175\alpha^{0.7023}-0.014\alpha^{0.7778}
 +0.0267\alpha^{0.8025}
 \non\\&&+0.0006\alpha^{0.8804}-0.1807\alpha^{1.417}
 +0.129\alpha^{1.444}-0.002\alpha^{1.802},
 \\
 C_5(\mu)&=&-0.0252\alpha^{-0.2977}+0.0227\alpha^{-0.1196}
 +0.0309\alpha^{0.4173}+0.0251\alpha^{0.7023}
 \non\\&&+0.0016\alpha^{0.7778}
 -0.0045\alpha^{0.8025}-0.0058\alpha^{0.8804}-0.0338\alpha^{1.417}
 \non\\&&+0.0348\alpha^{1.444}-0.026\alpha^{1.802},
 \\
 C_6(\mu)&=&0.019\alpha^{-0.2977}-0.0081\alpha^{-0.1196}
 +0.0821\alpha^{0.4173}-0.0624\alpha^{0.7023}
 \non\\&&-0.0048\alpha^{0.7778}
 -0.1241\alpha^{0.8025}-0.0038\alpha^{0.8804}+0.0979\alpha^{1.417}
 \non\\&&-0.1045\alpha^{1.444}-0.0486\alpha^{1.802},
 \ed
in terms of the running coupling constant
 \be
 \alpha=\alpha_s(\mu)={4\pi\over\beta_0\ln(\mu^2/\Lambda_{\overline{\rm
 MS}}^2)}\left[1
 -{\beta_1\over\beta_0^2}{\ln\ln(\mu^2/\Lambda_{\overline{\rm MS}}^2)
 \over\ln(\mu^2/\Lambda_{\overline{\rm MS}}^2)}\right],
 \ed
with the coefficients
 \be
 \beta_0={33-2f\over3},~~~\beta_1=102-{38\over3} N_f.
 \ed
For the active flavor number $N_f=3$ and the QCD scale
$\Lambda_{\overline{\rm MS}}=\Lambda_{\overline{\rm MS}}^{(3)}=375$
MeV, we have $\alpha_s(m_c)=0.407$. The initial conditions of the
Wilson coefficients at $\mu=m_c$ are given, in the NDR scheme, by
 \be
 C_1(m_c)&=&-0.43,~~C_2(m_c)=1.22,~~~
 \non\\
 C_3(m_c)&=&0.018,~~~C_4(m_c)=-0.046,
 \\
 C_5(m_c)&=&0.013,~~~C_6(m_c)=-0.044.\non
 \ed
Another useful set of values contains
 \be
 C_1(1\;{\rm GeV})&=&-0.51,~~C_2(1\;{\rm GeV})=1.27,~~~
 \non\\
 C_3(1\;{\rm GeV})&=&0.028,~~~C_4(1\;{\rm GeV})=-0.065,
 \\
 C_5(1\;{\rm GeV})&=&0.021,~~~C_6(1\;{\rm GeV})=-0.058,\non
 \ed
which are evaluated at the typical energy release in two-body nonleptonic $D$
meson decays.

The effective Wilson coefficient $C_{8g}^{\rm eff}$ for the magnetic
penguin operator is known only to the leading-logarithm accuracy.
Because of its smallness, we shall not
consider the evolution of $C_{8g}^{\rm eff}(\mu)$, but approximate
it by the constant
\be C_{8g}^{\rm eff}(m_c)=-0.11. \ed

\section{Chiral factor }\label{anomaly}

The chiral factor in Eq.~(\ref{chiral}) comes from the relation
between the matrix elements of the scalar (pseudoscalar) current
and of the vector (axial vector) current. Using the equations of
motion, we derive
 \be
& &\la P_2|\bar u \ga_5 q|0\ra={-i\partial^\mu \la P_2|\bar u
\ga_\mu\ga_5 q|0\ra\over m_u+m_q}
 =-i{f_{P_2}m_{P_2}^2\over m_u+m_q},\\
& &\la P_1|\bar q c|D\ra={-i\partial^\mu \la P_2|\bar q \ga_\mu
c|0\ra\over m_c}
 ={m_D^2-m_{P_1}^2\over m_c}F_0^{DP_1}(m_{P_2}^2),
 \ed
where the light quark mass has been neglected in the second
expression.

For the $\eta^{(\prime)}$ meson, the chiral factor receives an
additional contribution from the axial anomaly in the following
equation of motion \cite{kroll}
 \be
 \partial^\mu(\bar q \ga_\mu\ga_5 q)=2 m_q(\bar q i\ga_5
 q)+{\alpha_s\over4\pi}G\widetilde{G},\label{ax}
 \ed
in which $G$ represents the gluon field tensor, $\widetilde G$
is its dual, and $q$ denotes the quark fields $u$, $d$, and $s$. In
the present analysis only the $u$ quark contributes to the amplitude
$P_C$. We express the physical states $\eta$ and $\eta'$ as the
linear combinations of the flavor states $\eta_q$ and $\eta_s$ via a
mixing angle $\phi$,
\begin{equation}\label{etamixing}
\left(
  \begin{array}{cc}
    \eta\\ \eta'
  \end{array}
\right) = \left(
  \begin{array}{cc}
    \cos\phi & -\sin\phi \\
    \sin\phi & \cos\phi
  \end{array}
\right) \left(
  \begin{array}{c}
    \eta_q \\ \eta_s
  \end{array}
\right).
\end{equation}
The KLOE collaboration has extracted the value
$\phi=(40.4\pm0.6)^{\circ}$ recently \cite{KLOE}, which is
consistent with the phenomenological result in \cite{kroll}.

Equations~(\ref{ax}) and (\ref{etamixing}) then lead to the matrix
elements
 \be
 \la\eta|\bar u i\ga_5 u|0\ra
 &=&{1\over2m_u}\left({1\over\sqrt 2}m_{\eta}^2f_q\cos\phi
 -\cos\phi\la\eta_q|{\alpha_s\over4\pi}G\widetilde G|0\ra
 +\sin\phi\la\eta_s|{\alpha_s\over4\pi}G\widetilde G|0\ra\right),\\
 \la\eta'|\bar u i\ga_5 u|0\ra
&=&{1\over2m_u}\left({1\over\sqrt
2}m_{\eta'}^2f_q\sin\phi-\sin\phi\la\eta_q|{\alpha_s\over4\pi}G\widetilde
G|0\ra
 -\cos\phi\la\eta_s|{\alpha_s\over4\pi}G\widetilde G|0\ra\right),
 \ed
with the $\eta_q$ decay constant $f_q$. For the
expressions of the anomalies
 \be
 \la\eta_q|{\alpha_s\over4\pi}G\widetilde G|0\ra&=&\sqrt 2 f_q
 a^2,~~~~
 \la\eta_s|{\alpha_s\over4\pi}G\widetilde G|0\ra=y f_q a^2,
 \ed
the involved parameters have been determined to be $f_q=1.07f_\pi$,
$a^2=0.265$ GeV$^2$, and $y=0.81$ \cite{kroll}.

\section{Scalar matrix elements}\label{me}

In this Appendix we fix the scalar matrix element $\la P_1P_2|\bar
q_1q_2|0\ra$ appearing in Sec.~III C, which is dominated by the
contribution from the lowest scalar resonance \cite{Choi:1998yx},
 \be\label{sme}
 \la P_1P_2|\bar q_1 q_2|0\ra=\la P_1P_2|S\ra\la S|\bar q_1
 q_2|0\ra=g_{S}B_{S}(q^2)m_{S}\bar f_{S},
 \ed
as shown in Eq.~(\ref{bs}). The scalar decay constant $\bar f_{S}$
is defined via $\la S|\bar q_2 q_1|0\ra=m_S\bar f_S$
\cite{Cheng:2005nb}. Many scalar mesons have been observed. It is
still controversial whether the lightest scalar nonet with the mass
smaller than or close to 1 GeV are primarily the four-quark or
two-quark bound states. Since $\kappa$, $a_0(980)$, and $f_0(980)$
may be the four-quark states, we consider the scalar resonances
$a_0(1450)$ for the $\pi\eta$, $\pi\eta'$, and $(KK)^\pm$ final
states, $f_0(1370)$, $f_0(1500)$, and $f_0(1710)$ for the
$\pi^0\pi^0$, $\pi^+\pi^-$, $K^0\bar K^0$, $K^+K^-$, and
$\eta\eta^{(\prime)}$ final states, and $K_0^*(1430)$ for the $K\pi$
and $K\eta^{(\prime)}$ final states due to the symmetries of strong
interaction. $f_0(1370)$, $f_0(1500)$, and $f_0(1710)$
are the mixtures of $n\bar n\equiv(u\bar u+d\bar d)/\sqrt 2$, $s\bar s$
and glueball, but there still exists controversy on the components of
each scalar meson. We employ the mixing matrix obtained in
\cite{Cheng:2006hu} in this work. Note that there are no $s$-wave
isospin-1 resonances for the $\pi^\pm\pi^0$ system under the isospin
symmetry.

We derive $\la K\pi|\bar s u|0\ra$ as an example.
Inserting $K_0^*(1430)$ as the intermediate resonance,
the matrix element is written as
 \be
 \la K\pi|\bar s u|0\ra=\la K\pi|K_0^*(1430)\ra\la K_0^*(1430)|\bar s
 u|0\ra=g_{K_0^*K\pi}B_{K_0^*}(q^2)m_{K_0^*}\bar f_{K_0^*},
 \ed
where the effective strong coupling constant $g_{K_0^*K\pi}=3.8$ GeV
can be obtained directly from the measured $K_0^*(1430)\to K\pi$
decay \cite{Fajfer:1999hh}, and $B_{K_0^*}$ is the $K_0^*(1430)$
propagator in the Breit-Wigner form \cite{Choi:1998yx}
 \be
 B_{K_0^*}(q^2)={1\over
 q^2-m_{K_0^*}^2+im_{K_0^*}\Gamma_{K_0^*}(q^2)},
 \ed
with the momentum-dependent width
 \be
 \Gamma_{K_0^*}(q^2)&=&\Gamma_{K_0^*}D_{K_0^*}(q^2),\non\\
 D_{K_0^*}(q^2)&=&{m_{K_0^*}\over\sqrt{q^2}}{P_K(q^2)\over
 P_K(m_{K_0^*}^2)},\non
 \\
 P_K(q^2)&=&{1\over2\sqrt{q^2}}\lambda^{1/2}(q^2,m_\pi^2,m_K^2),
 \non\\
 \lambda(x,y,z)&=&x^2+y^2+z^2-2xy-2yz-2xz.
 \ed
The values of the mass $m_{K_0^*}$ and the decay width
$\Gamma_{K_0^*}$ for the scalar resonance
$K^*_0(1430)$ are taken from the Particle Data Group 2010
\cite{PDG}.

The relevant scalar decay constants are given, in QCD sum rules, by
\cite{Cheng:2005nb}
 \be
 \bar f_{n\bar n}(1\; \rm GeV)&=&(460\pm50)\;\rm MeV,
 \non\\
 \bar f_{s\bar s}(1\;\rm GeV)&=&(490\pm50)\;\rm MeV,
 \non\\
 \bar f_{K_0^*(1430)}(1\;\rm GeV)&=&(445\pm50)\;\rm MeV,
 \ed
where $\bar f_{n\bar n}$ is the scalar decay constant for
$a_0(1450)$ and for the $n\bar n$ components of the three iso-singlet
scalars, and $\bar f_{s\bar s}$ for the $s\bar s$ components of the three
iso-singlet scalars. We assume that the quark, but not glueball, components
of scalars dominate the $D$ meson decays. With
lack of experimental data, the effective coupling constant $g_S$ in
Eq.~(\ref{sme}) is approximated by $g_S=g_{K_0^*K\pi}$ in the flavor
$SU(3)$ limit.

\end{appendix}

\end{document}